\documentclass[11pt, oneside]{article}   	
\usepackage{geometry}                		
\usepackage{graphicx}				
\usepackage{amssymb}
\newtheorem{theorem}{Theorem}
\newtheorem{lemma}[theorem]{Lemma}
\title{Singularities of  scattering matrix}

\begin{document}
\author {  Albert Schwarz\\ Department of Mathematics\\ 
University of 
California \\ Davis, CA 95616, USA,\\ schwarz @math.ucdavis.edu}

\maketitle
\begin {abstract} 
Our main result is the analysis of singularities of integrands of integrals representing matrix elements of scattering matrix and inclusive scattering matrix in perturbation theory. These results are proven for any quantum field theory in any dimension. 
\end {abstract}

\section {Introduction}
The primary goal of the present paper was the analysis of the inclusive scattering matrix.  This analysis led to some results also for the conventional scattering matrix. Namely, in both cases, we get some information about singularities of integrands of corresponding integrals (under certain assumptions). We propose a procedure that allows us to analyze singularities of integrands for inclusive scattering matrix by induction.  This procedure gives information about the conventional scattering matrix, but it cannot be formulated without using the inclusive scattering matrix.
 
 It was proven in \cite {A} and other papers that scattering amplitudes for N=4 super Yang-Mills and in some other theories can be expressed in terms of on-shell diagrams. We found that in these cases inclusive scattering matrix also can be expressed this way.
 
 Many recent papers are based on the idea that it is possible
 to formulate quantum field theory "on shell"  ( in terms of particles instead of fields).   This idea led to very interesting (but still incomplete) results for four-dimensional massless theories. Our results can be applied in any dimension to any theory. They are incomplete, however, they strongly support this idea.

Our main tools are generalized Green functions (GGreen functions). These functions appear naturally in the formalism of $L$-functionals and in Keldysh formalism. (See \cite {MO}, \cite {SCI}  for a  review of the formalism of $L$-functionals, suggested in \cite {SCH},  and\cite {UNI}, \cite {CU}, \cite {K} for the review of Keldysh formalism.)

One can define inclusive $S$-matrix as  on-shell GGreen function.  In the case when the theory has particle interpretation inclusive cross-sections can be expressed linearly
 in terms of matrix  entries of inclusive $S$-matrix \cite {MO}, \cite {T}, \cite {SCI}, \cite {SCA}.

 Notice that it is possible to modify our considerations in a way that allows us to analyze the inclusive scattering matrix of quasiparticles (elementary excitations of equilibrium states or, more generally, translation-invariant stationary state). In this case, the conventional scattering matrix does not have any physical meaning, but one can consider the inclusive scattering matrix neglecting the instability of quasiparticles.   

\section {GGreen functions}
 One can define generalized Green functions (GGreen functions) in the state $\omega$ by the following formula where $B$ is an  observable:

 \begin{equation}\label {GG}
 G_n^S=\omega (MN)
 \end {equation}
 where $$M= T^{opp}( B^*({\bf x}_{i_1}, t_{i_1})\dots B^*({\bf x}_{i_s},t_{i_s}))$$ stands for antichronological product (times increasing)
 and
 $$N=T(B({\bf x}_{j_1},t_{j_1})\dots B({\bf x}_{j_{n-s}},t_{j_{n-s}}))$$ stands for chronological product (times decreasing). ( We consider a state as a positive linear functional on the algebra of observables $\cal A$. Translations act as automorphisms of $\cal A$; spatial translation by $\bf x$ and time translation by $t$ transform $B$ into $B({\bf x},t).$ )
 
 The GGreen function depends on the choice of the subset $S$ of the set $\{1,\cdots n\}$, $j_k$ belong to $S$, $i_l$ belong to the complement of $S$. If $S$ coincides with  $\{1,\cdots n\}$ the GGreen function will be denoted by $G_n^+$; it coincides with the conventional Green function. If $S$ is empty the GGreen function will be denoted by $G_n^-$; it is a complex conjugate to $G_n^+.$
 
 The observable $B$ can depend on a discrete variable (spin, polarization, etc), then the GGreen function also depends on discrete variables.
 
 Instead of the notation $G_n^S$ we use the notation 
 $G_n({\bf x}_1,t_1,\epsilon_1,..., {\bf x}_n, t_n,\epsilon_n)$  where 
 $\epsilon _i=\pm 1.$  The set $S$ is identified with the set of all $i$'s obeying $\epsilon_i=+1.$
 As usual after Fourier transform we obtain GGreen functions in $({\bf p},t)$- and $ ({\bf p}, \omega)$-representations. Notice that in  $ ({\bf p}, \omega)$-representation  the complex conjugate of $G_n({\bf p}_1, \omega_1, \epsilon_1,..., {\bf p}_n, \omega_n, \epsilon_n)$ is equal to $G_n(-{\bf p}_1, -\omega_1,- \epsilon_1,..., -{\bf p}_n, -\omega_n, -\epsilon_n)$  (to   $G_n(-{\bf p}^*_1, -\omega^*_1,- \epsilon_1,..., -{\bf p}^*_n, -\omega^*_n, -\epsilon_n)$  if we consider analytic extension to complex arguments).
 
The inclusive cross-section can be expressed in terms of amputated on-shell GGreen functions in  $ ({\bf p}, \omega)$-representation. To obtain this cross-section we exclude the $\delta$-function corresponding to energy-momentum conservation from the GGreen function;  in the remaining expression we take $+$ variables equal to $- $variables. More precisely,  if we are interested in the collision of two particles with momenta ${q}_1, {\ q}_2$ 
producing particles with momenta ${ p}_1, ...,{ p}_n$ plus some unspecified particles we should consider amputated
GGreen function $G_{2n+4}$ depending on  $n+2$ on-shell variables  ${ p}_1, ...,{ p}_n, { q}_1, { q}_2$  with $\epsilon =+1$ and $n+2$ on-shell variables ${ p}'_1, ...,{p}'_n, {q}'_1, { q}'_2$  with $\epsilon=-1.$ We should exclude the $\delta$-function  from $G_{2n+4}$ and take $p'_i\to p_i, q'_j\to q_j.$ 
 
 We are mostly interested in connected amputated on-shell GGreen functions in $ ({\bf p}, \omega)$-representation.   We denote these functions by
 $\hat G_n({ p}_1,\epsilon_1,...,{ p}_n, \epsilon_n)$ (For simplicity we assume that there exists only one type of particles with dispersion law $\omega(\bf p)$; to get on-shell GGreen function we take $\omega_i=\omega({\bf p}_i)$ in amputated GGreen function.) 
 
 One can prove (see, for example \cite {WEL}) that the sum of all functions $\hat G_n$ vanishes:
 \begin {equation}\label{PM}
 \sum_{\epsilon _i=\pm 1}\hat G_n({ p}_1,\epsilon_1,...,{ p}_n, \epsilon_n)=0
 \end {equation}

 This relation allows us to express the function $\hat G_n^++\hat G_n^-$  
  in terms of other on-shell GGreen functions. 
  
  The relation (\ref {PM}) is closely related to the unitarity of the scattering matrix and to Cutkosky's cutting rules.
  
  In the formalism of $L$-functionals we work with linear functionals on Weyl algebra ( an associative algebra with involution with generators obeying canonical commutation relations). Such a functional (denoted by $L_K$)  corresponds to every trace class operator $K$ in any representation of CCR: if $A$ is an element of Weyl algebra then $L_K(A)=Tr AK.$ A  field $\phi$ generates two operators on the 
  space $\cal L$ of linear functionals on Weyl algebra; one of them (denoted by $\phi_+$) corresponds to the multiplication of $K$ by $\phi$ from the left, the second one (denoted by $\phi_-$) corresponds to the multiplication of $K$ by $\phi^*$ from the right. If $K$ is a density matrix $L_K$ is a physical $L$-functional describing a state of our system.
  
  Applying a chronological product of operators $\phi_+({\bf x}_j,t_j)$ and operators $\phi_-({\bf x}_i,t_i)$ to the state $\omega$ we obtain a linear functional on the Weyl algebra. It is easy to check that by calculating the value of this functional on the unit element of Weyl algebra we obtain the GGreen function. This remark allows us to construct in the standard way the diagram techniques for the calculation of GGreen functions. The same techniques appear in Keldysh formalism.
 
The diagram techniques that allow us to calculate GGreen functions in the framework of perturbation theory are very similar to the techniques for conventional Green functions.
 We need very limited information about the diagrams calculating GGreen functions.
First of all, if we had $n$ fields in the original Lagrangian or Hamiltonian then in the diagrams  for GGreen functions we should have $2n$ fields (+ and - fields) 
These diagrams have vertices of two types (+ and - vertices). The crucial property of propagators: the propagators connecting vertices of different types (+- propagators and -+ propagators) are on shell.


For example for scalar field $$G_{+-}(p)=\hbar 2\pi\theta (-\omega) \delta (p^2-m^2),G_{-+}(p)=\hbar 2\pi \theta (\omega) \delta (p^2-m^2), $$\\$$G_{++}(p)=\hbar\frac {i}{p^2-m^2+i0}, G_{--}=\hbar\frac {-i}{p^2-m^2-i0},$$
where $p=(\omega, \bf p)$ denotes the energy-momentum vector.
The verices contain a factor $\hbar ^{-1}.$

(We assume here that the GGreen functions correspond to
 the ground state,  although our considerations can be applied in more general situations).

{\it In what follows we are using renormalized diagram technique where the propagators contain physical dispersion law and are properly normalized.  Our considerations do not depend on Lorentz invariance and locality; therefore we do not care about divergences.}

\section{Semiclassical approximation}
Notice that one can get a little bit different diagram technique taking different basis in the space of fields (Keldysh basis or, in different terminology, physical basis); this technique was used in quantum field theory in \cite {TL}. In this basis we replace $\phi_+,\phi_-$ with 
$$\phi^r=\frac 1 2 (\phi_++\phi_-), \phi^a=\hbar^{-1}(\phi_+-\phi_-).$$
(Alternative notations are $\phi^r=\phi_{cl},\phi^a=\phi_{qu}$ where $cl$ stands for classical and $qu$ for quantum.) 
One can define GGreen functions in terms of these fields; in coordinate representations they depend on variables $({\bf x}_1,t_1,\sigma_1,..., {\bf x}_n, t_n,\sigma_n)$ where $\sigma_i=r$ or $\sigma_i=a.$ These functions  are linear combinations of  functions   $G_n({\bf x}_1,t_1,\epsilon_1,..., {\bf x}_n, t_n,\epsilon_n)$  with constant coefficients. The function with all $\sigma_i=a$ is equal to zero; this statement is equivalent to (\ref {PM}). As usual, we can define GGreen functions in ${\bf p},t$ and in ${\bf p},\varepsilon$ representations. The inclusive scattering matrix can be expressed in terms of GGreen function  $G_n(p_1,\sigma_1,...,p_n,\sigma_n)=G_n({\bf p}_1,\varepsilon_1,\sigma_1,..., {\bf p}_n, \varepsilon_n,\sigma_n)$  on shell. The diagram technique in Keldysh basis is very similar to the technique described in Section 2. The propagator can be regarded as $2\times 2$ matrix where the diagonal entry $G_{aa}$ vanishes and the diagonal entry $G_{rr}$ is on shell. 

 For scalar field
the propagators in Keldysh basis are
$$G^{rr}= \hbar\pi \delta (p^2-m^2), G^{aa}=0, G^{ra}=G^{ar} =\frac{1}{p^2-m^2+i\omega0}$$ It is easy to check that the vertices with  indices $rr...r$ vanish. 
A vertex having $k$ indices of type $a$ contains a factor $\hbar ^{k-1}.$  It follows that the representation of inclusive scattering matrix by diagrams in Keldysh basis contains only non-negative powers of $\hbar$, hence it gives a decomposition of this matrix in Taylor series with respect to $\hbar$  (see \cite {INC} for more detail). In particular, the limit of inclusive scattering matrix as $\hbar\to 0$ is represented as a sum of diagrams where all propagators are of the form $G^{ra}, G^{ar}$ and all vertices have only one index of type $a.$

\section {On-shell diagrams}

In on-shell diagrams edges are oriented; as usual, every  edge should carry momentum and every vertex should contain a delta function expressing the conservation of momentum. The  propagators should be on shell (they should have the form 
$M(p)\delta (\omega-\omega({\bf p}))$  or $M(p)\delta (\omega+\omega({\bf p}))$ )and the momenta of all external vertices should be on shell.

All on-shell diagrams we consider have vertices of two types (+ and - vertices).

It is obvious that any {\it on-shell GGreen function can be represented as a sum of on-shell diagrams with propagators } $G_{+-}(p),  G_{-+}(p)$ {\it  and with  $\hat G_n^+$ , $\hat G_n^-$ as vertices}.  (Replacing   vertices $\hat G_n^+$ and $\hat G_n^-$ with diagrams representing these functions we obtain  diagrams for an on-shell GGreen function.)

This statement allows us to express all loop level $l$ on-shell GGreen functions in terms of functions  $\hat G_n^+$ and $\hat G_n^-$ at the loop levels $\leq  l.$  {\it If the conventional on-shell Green functions   $\hat G_n^+$  can be expressed in terms of on-shell data the same is true for on-shell GGreen functions}. In particular, {\it if on-shell Green functions are represented by on-shell diagrams on-shell GGreen functions also are represented by on-shell diagrams}.

 BCFW recursion allows us to express scattering amplitudes in terms of simpler scattering amplitudes (at least in the case of vanishing boundary contribution). Combining this fact with the above statements we obtain that for all theories we can express tree-level on-shell GGreen functions in terms of on- shell data  (and in the case of gauge theories in terms of on-shell diagrams).
 
 For N=4 SUSY Yang-Mills theory one can express all on-shell  Green functions, hence all on-shell GGreen functions in terms of on-shell diagrams.
\section {Partial summation}
 
  \begin {lemma} \label {5}
   The sum of all diagrams for  GGreen function
  $\hat G_n (p_1,\epsilon_1, ..., p_n,\epsilon_n)$ containing a  $++$ edge  separating
  $p_1,...,p_{k}$ from  $p_{k+1},...,p_n$ is equal to
  $$\int dq G_{++}(q)\hat G_k(p_1,\epsilon_1,...,p_k, \epsilon_k , q,+1)
  \hat G_{n-k}(p_{k+1},\epsilon_{k+1},..., p_n, \epsilon_n, -q,+1)$$

 \end {lemma}
 
 The lemma is an obvious generalization of well-known statement for conventional Green functions.

 Let us consider a diagram for a connected on-shell GGreen function with $n$ external vertices. Let us remove the edge  corresponding  to +- propagator. Denote the remaining part by $A$; we  add to this part two external vertices ($+$ vertex and $-$ vertex)  with momenta $q,-q$ where $q$ is an on-shell momentum of the removed edge. The contribution of $A$ to the GGreen function will be denoted by $A_{n+2} (p_1, ..., p_n, q,-q)$.   
 It is easy to prove the following 
 
 \begin {lemma} \label {1}

 We can get the contribution of the original diagram to the GGreen function multiplying
 $A_{n+2}$ by  $G_{+-}(q)$ and integrating over $q.$.
\end {lemma}

 The set $A$ can be connected or disconnected (the removed +- edge can be non-separating or separating).
 \begin {lemma} \label {2}
  The sum of all diagrams for on-shell GGreen function
  $\hat G_n (p_1,\epsilon_1, ..., p_n,\epsilon_n)$ containing a $+-$ edge separating
  $p_1,...,p_k$ from $p_{k+1},...,p_n$ is equal to
  $$\int dq G_{+-}(q)\hat G_{k+1}(p_1,\epsilon_1,...,p_k, \epsilon_k, q, +1)
  \hat G_{n-k+1}(p_{k+1},\epsilon_{k+1},..., p_n, \epsilon_n, -q,-1)$$
  \end {lemma} 
  This lemma is similar to Lemma \ref {5}, the proof is the same.

 \begin {lemma} \label {3}
   The sum of all diagrams for on-shell GGreen function
  $\hat G_n (p_1,\epsilon_1, ..., p_n,\epsilon_n)$ containing a non-separating $+-$ edge is equal to
   $$\int dq G_{+-}(q)\hat G_{n+2}(p_1,\epsilon_1,...,p_n,\epsilon_n,q,+1, -q,-1).$$
   \end {lemma}

To prove  Lemma  \ref {3} we start with any diagram for the function $\hat G_{n+2}(p_1,\epsilon_1,...,p_n,\epsilon_n,q,+1, q,-1).$  Connecting the vertices with momenta $q,-q$ with $+-$ edge obtain a diagram for   $\hat G_n (p_1,\epsilon_1, ..., p_n,\epsilon_n)$ with non-separating edge. (Recall that  all diagrams for the functions $\hat G$ are connected.) All diagrams with non-separating $+-$ edge can be obtained this way. Now we can apply  Lemma \ref {1}.



\section {Singularities}

The statements of the preceding section can be used to obtain information about singularities of the (inclusive) scattering matrix.
In many cases, it is useful to work with integrands of integrals expressing the (inclusive) scattering matrix (see \cite {AT}, \cite {A}, etc).  It was shown in these papers that in the case of planar N=4 SUSY Yang-Mills the singularities of these integrands can be used to calculate the integrands. Later this statement was generalized.

We assume that the singularities (or at least leading singularities)   of partial sums are also singularities of GGreen functions we are studying.  Moreover, we assume that a similar statement is true for corresponding integrands. Of course, these statements are not necessarily correct, the singularities of partial sums can cancel against the singularities of remaining summands.

 We represent a GGreen function as a sum of Feynman diagrams.  Feynman diagram in every order of perturbation theory is an integral over internal momenta; we represent the GGreen function in given order of perturbation theory 
as an integral of the sum of the integrands of individual diagrams.  In what follows talking about GGreen function we have in mind the connected GGreen function in a fixed order of perturbation theory.
The integrand of an integral representing the GGreen function is not well defined; for example, in gauge theories it depends on the choice of gauge condition.  Talking about an integrand we have in mind one of the integrands. One can hope that our statements are true for all (or almost all) integrands.

We denote the integrand of GGreen function $\hat G$ by $\hat g$. Considering the $l$-loop contribution we are using the notation ${\hat G} ^l$ for the GGreen function and ${\hat g}^l$ for the integrand. The integrand can be considered as a differential form on the space of internal momenta. (Sometimes it is convenient to  consider it as a differential form on the space of internal and external momenta.) Notice that the momenta are not independent  (or, equivalently, the coefficients of the differential form contain delta functions coming from conservation laws).

In the conditions of Lemma \ref {2}  the expression 
  \begin{equation}\label {GG}
 \int  dqG_{+-}(q) \sum _{l_1+l_2=l} \hat g_k^{l_1}(p_1,\epsilon_1,...,p_k, \epsilon_k ,q,+1)
  \hat g_{n-k}^{l_2}(p_{k+1},\epsilon_{k+1},..., p_n, \epsilon_n, -q,-1)
  \end{equation}
  is  a partial integrand for   on-shell GGreen function $\hat G_n^l (p_1,\epsilon_1, ..., p_n,\epsilon_n)$.

It follows from  Lemma \ref {3} 
\begin{equation}\label{GGG}\int dq G_{+-}(q)\hat g^{l-1}_{n+2}(p_1,\epsilon_1,...,p_n,\epsilon_n,q,+1, -q,-1)
\end{equation}
is a partial integrand for the same GGreen function.

One can use these statements to describe an inductive procedure that allows us to calculate the singularities of $\hat g_n (p_1,\epsilon_1, ..., p_n,\epsilon_n)$. This procedure is  similar to the procedure described in \cite {AT}, \cite {A} (see the formula (2.25) in \cite {A}).

The relations (\ref {GG}) and (\ref {GGG}) do not describe singularities of integrands $\hat g^+_n$  of conventional Green functions   $\hat G^+_n$. Some information about these singularities can be obtained from  Lemma \ref {5}.  We can use (\ref {PM}) to get additional information.

Using (\ref {PM}) we derive from Lemma \ref {3} or from the relation (\ref {GGG}) that
\begin{lemma}\label{6}
$-  \int dq G_{+-}(q)\sum' \hat g_{n+2}(p_1,\epsilon_1,...,p_n,\epsilon_n,q,+1, -q,-1)$
is a partial integrand for  the exptression  $T_n(p_1,...,p_n)=(\hat G^+_n+\hat G^-_n)(p_1,...,p_n)$.
(The sign $\sum'$ denotes summation over all $\epsilon _i=\pm 1$ except 
$\epsilon_1=...=\epsilon_n=1$ and $\epsilon_1=...=\epsilon_n=-1.$). 
\end{lemma}
 This statement allows us to obtain information about the singularities of the integrand of $T_n$ and therefore about the singularities of the integrands of $ \hat G^+_n$ and $\hat G^-_n.$

Additional information about singularities of $T_n$ can be obtained in the same way from (\ref {PM}) and
(\ref {GG}).

\begin {lemma}
$-\sum'\int  dqG_{+-}(q) \sum _{l_1+l_2=l} \hat g_k^{l_1}(p_1,\epsilon_1,...,p_k, \epsilon_k ,q,+1)
  \hat g_{n-k}^{l_2}(p_{k+1},\epsilon_{k+1},..., p_n, \epsilon_n, -q,-1)$
  is a partial integrand for  the exptression  $T_n(p_1,...,p_n)=(\hat G^+_n+\hat G^-_n)(p_1,...,p_n)$.
(The sign $\sum'$ denotes summation over all $\epsilon _i=\pm 1$ except 
$\epsilon_1=...=\epsilon_n=1$ and $\epsilon_1=...=\epsilon_n=-1.$). 
\end{lemma}
Now we can suggest an inductive procedure that allows us to obtain information about the singularities of integrands for on-shell GGreen functions. Calculating  $\hat g_n$  we denote by  $\hat g_n^{l,s}$ the contribution  of the diagrams with the number of loops $\leq l$ and the number of $+-$ ans $-+$ propagators $\leq s.$
Using Lemma 5 we obtain information about singularities of $\hat g_n^{l,s}$ from the information
about singularities of $\hat g_{n+2}^{l-1, s-1}.$ Using (\ref {GG}) we obtain information about singularities of $\hat g_n^{l,s}$ from the information about singularities of $\hat g_n^{l,s-1}.$ As a result we get information about the singularities of $\hat g_n^{l,s}$ from the information about singularities of $\hat g_{n+2s}^{l,0.}$

To calculate singularities of $\hat g_n^{l,0}$ we should know the singulaties  of $l$-loop contribution  to  $\hat g_k$  for $k\leq n.$
The information about these singularities can be obtained from Lemma 1 and Lemma 5 if we know the singularities of $\hat g_k^{l-1,s}$.

 In Lorentz-invariant theories
  the integrands are rational functions. ( We consider $\delta (x)$ as a rational function of $x$ because it can be expressed as a linear combination of two fractions $\frac {1}{x\pm i0}.$)  Therefore one can hope that the above statements describe all singularities. (In other theories  additional singularities come from the singularities of the dispersion law.)
  
  \section {Relation to positive Grassmannian}
  
   It seems that deep relations with cluster algebras and positive Grassmannian discovered in \cite {A} for N=4 SUSY Yang-Mills are very general. They are based first of all on the consideration of on-shell diagrams with two types of vertices (bicolored graphs in mathematical terminology). Such diagrams appear also in our approach (however the origin of the two types of vertices is completely different).  To relate these diagrams to positive Grassmannian and cluster algebras one notices that a planar diagram of this kind  (planar bicolored graph=plabic graph) specifies a positroid ( a cell in a  cell decomposition of positive Grassmannian)  and a cluster in the corresponding cluster algebra.  Different plabic graphs specify the same positroid if they are related by a sequence of local moves (cluster transformations in the language of cluster algebras).  The relation to physics was derived in \cite {A}  from the remark that these moves do not change the amplitude corresponding to an on-shell diagram. If this remark could be generalized to our situation we would be able to use the techniques of positive Grassmannian.

    { \bf Acknowledgments} I am indebted to J. Bourjaily, A. Konechny,  M. Movshev, A. Rosly, and  J. Trnka for useful discussions.

\end{document}